\pdfoutput=1        

\documentclass[sigplan,10pt,nonacm]{acmart}         
\makeatletter                   
\def\mdseries@tt{m}             
\makeatother        

\makeatletter
\renewcommand\@formatdoi[1]{\ignorespaces}
\makeatother

\usepackage[export]{adjustbox}
\usepackage{listings}
    
\usepackage{color}                                                                   
\usepackage{xcolor}                                            
\usepackage{colortbl}                       
\usepackage{changepage} 
                                                                                     
\usepackage{url}                                                                     
\hypersetup{hidelinks=false,colorlinks=true,linkcolor=blue,urlcolor=blue,citecolor=blue,anchorcolor=blue} 
\usepackage{breakurl}                                                                

\usepackage{tabularx}
\usepackage{booktabs} 

\usepackage{cleveref}
\crefformat{section}{\S#2#1#3}
\crefformat{subsection}{\S#2#1#3}
\crefformat{subsubsection}{\S#2#1#3}
\crefrangeformat{section}{\S\S#3#1#4 to~#5#2#6}
\crefmultiformat{section}{\S\S#2#1#3}{ and~#2#1#3}{, #2#1#3}{ and~#2#1#3}

\usepackage{microtype}
\usepackage[utf8]{inputenc}                                                          

\usepackage[iso,american,cleanlook]{isodate}    

\usepackage{rviewport}                                                               

\usepackage{tikz} 
\usepackage[inline]{enumitem} 

\newcommand\Invisible[1]{                                                            
  \marginpar{\color{white}{\fontsize{.5}{.5}\selectfont #1 }}                        
}

\newcommand{\Exclude}[1]{}

\newcommand\Boldly[1]{\vspace{0.5 \baselineskip} \noindent \textbf{$\blacktriangleright$} \textbf{#1} \noindent}
\newcommand\BoldSection[1]{\vspace{0.5 \baselineskip} \noindent \textbf{#1} \noindent}

\definecolor{Gray95}{gray}{0.95}
\definecolor{forestgreen}{rgb}{0.13, 0.55, 0.13}
                                                                                     
\usepackage[scaled]{beramono} 
\usepackage[T1]{fontenc}

\newcommand{\AtFoot}[1]{\let\thefootnote\relax\footnotetext{{#1}}}

\acmConference[]{}{}{}
\acmYear{}
\acmISBN{} 
\acmDOI{} 
\startPage{1}

\usepackage{booktabs}   
\usepackage{subcaption} 

\newcommand{\orcidicon}[1]{\href{https://orcid.org/#1}{\includegraphics[scale=0.06]{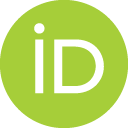}}}

\begin{document}

\title[]{Fissile Locks} 


\author{Dave Dice \orcidicon{0000-0001-9164-7747}}
\orcid{0000-0001-9164-7747}             
\affiliation{
  \institution{Oracle Labs}             
}

\email{first.last@oracle.com}            

\author{Alex Kogan \orcidicon{0000-0002-4419-4340}} 
\orcid{0000-0002-4419-4340} 
\affiliation{
  \institution{Oracle Labs}             
}
\email{first.last@oracle.com}          



\begin{abstract}

Classic test-and-test (TS) mutual exclusion locks are simple,
and enjoy high performance and low latency of ownership transfer under
light or no contention. They do not, however, scale gracefully under high
contention and do not provide any admission order guarantees.
Such concerns led to the development of scalable queue-based locks,
such as a recent \emph{Compact NUMA-aware} (CNA) lock, 
a variant of another popular queue-based \emph{MCS} lock. 
CNA scales well under load and provides certain admission guarantees, 
but has more complicated lock handover operations than TS and incurs
higher latencies at low contention.

We propose \underline{\textbf{Fissile}} locks, which capture the most desirable properties
of both TS and CNA. A Fissile lock consists of two underlying locks: a
TS lock, which serves as a fast path, and a CNA lock, which serves as a slow path.
The key feature of Fissile locks is the ability of threads on the fast path to bypass
threads enqueued on the slow path, and acquire the lock with less overhead than CNA. 
Bypass is bounded (by a tunable parameter) to 
avoid starvation and ensure long-term fairness. The result is a highly scalable 
NUMA-aware lock with progress guarantees that performs like TS at low contention
and like CNA at high contention.

\Invisible{AKA: Fissile Locks; Tetris Locks; bifurcated locks; passing lane} 
\Invisible{Tetris -- on-line stream packing problem with limited visibility} 
\Invisible{MCS provides FIFO admission and CNA provides bounded long-term fairness guarantees.} 
\Invisible{Notably CNA} 
\Invisible{Unlike classic CNA, our variant in Fissile reorganizes the chain of waiting threads early, 
immediately after acquiring the CNA lock.  As such, reorganization runs outside the
TS critical section and potentially allows overlap with execution of the critical section.} 
\Invisible{The thread that has acquired the TS lock is the owner of the compound Fissile lock.} 

\end{abstract}




\maketitle

\thispagestyle{fancy}

\Invisible{
  *  High contention; high traffic; arrival rate; intensity; transits; 
  *  Seize; barge; pounce; bypass; cut in line; jump line; snatch; 
  *  Rapacious locks
  *  Price-of-anarchy; tragedy-of-the-commons;
} 

\section{Introduction}


\BoldSection{TS: }
Test-and-test locks (TS) \cite{tpds90-Anderson} are compact --
consisting of a single \emph{lock word} -- simple,
and provide excellent latency under light or no contention.  
They fail to scale, however, as contention increases.  

Acquiring threads simply busy-wait, or \emph{spin} attempting to change the
lock word state from \emph{unlocked} to \emph{locked} with an atomic read-modify-write
instruction, such as \texttt{compare-\allowbreak{}and-\allowbreak{}swap} (CAS) or \texttt{exchange} (SWAP).  
If the atomic operation was successful, then 
the thread has acquired the lock and may enter the critical section.  
Releasing the lock requires only a simple store to set the state to unlocked.  
So-called ``polite'' test-and-test-and-set locks (TTS), a variation on TS, 
first fetch the lock value and only attempt the atomic instruction if the lock 
was observed to be not held.  
That is, acquiring threads busy-wait until the lock is clear, at which point they execute 
an atomic instruction to try to gain ownership.  
TTS acts to avoid unnecessary write invalidation arising from failed atomic operations. 
Simple ``impolite'' TS locks do not bother to first load the value, so each
\emph{probe} of the lock causes writing via the atomic instruction.  
TS and TTS locks are usually augmented with back-off -- delays between probes --
to moderate contention. 

\Invisible{Probe the lock} 

\Invisible{In our descriptions we will assume a sequentially consistent memory model 
and not consider the need for memory fence or barrier instructions.} 

\BoldSection{MCS: } 
The \emph{MCS lock} \cite{tocs91-MellorCrummey}, is the usual alternative to simple
test-and-set-based locks, performing better under high contention, but
also having a more complex path and often lagging behind simple locks under no or light contention.
In MCS, arriving threads use an atomic operation to append an element
to the tail of a linked list of waiting threads,
and then busy wait on a field within that element, avoiding global spinning as found in TS.
The list forms a queue of waiting threads.  
The lock's tail variable is explicit and the head -- the current owner --
is implicit. When the owner releases the lock it reclaims the element it
originally enqueued and sets the flag in the next element, passing ownership.
To convey ownership, the MCS unlock operator must identify the successor, if any, and then
store to the location where the successor busy waits.
The list forms a multiple-producer-single-consumer (MPSC) queue where any thread can enqueue but
only the current owner can dequeue itself and pass ownership.  
The handover path is longer than that of TS locks and accesses more distinct
shared locations.  

MCS uses so-called local waiting where at most one thread is waiting on a given
location at any one time.  As such, an unlock operation will normally
need to invalidate just one cache line -- the line underlying the flag where the successor busy waits --
in one remote cache.
Under contention, the unlock operator must fetch the address of the successor
element from its own element, and then store into the flag in the successor's element,
accessing two distinct cache lines, and incurring a dependent memory access to reach the successor.
Absent contention, the unlock operator uses an atomic compare-and-swap (CAS) 
operator to try to detach the owner's element and set the tail variable to \texttt{null}. 

MCS locks provide strict FIFO order.  They are also compact, with the lock
body requiring just a pointer to the tail of the chain of queue elements.  

One MCS queue element instance is required for each lock a thread currently holds, and
an additional queue element is required while a thread is waiting on a lock.
Queue elements can not be shared concurrently and can appear on at most one queue
-- be associated with at most one lock -- at a given time.
The standard POSIX \texttt{pthread\_mutex\_lock} and \texttt{pthread\_\allowbreak{}mutex\_\allowbreak{}unlock} 
operators do not require scoped or lexically balanced locking.  
As such, queue element can not be allocated on stack.   
Instead, MCS implementations that expose a standard POSIX interface will typically allocate elements 
from thread-local free lists, populated on 
demand \footnote{We note that the MCS ``K42'' variant \cite{K42,Scott2013} allows queue elements to 
be allocated on stack -- they are required only while a thread waits -- 
but at the cost of a longer path with more accesses to shared locations.}. 

\Invisible{MCS requires the address of queue element inserted by the owner to 
be passed to the corresponding unlock operator, where it will be used to identify 
a successor, if any.} 

The standard POSIX interface does not provide any means to pass information 
from a lock operation to the corresponding unlock operator.  As such, the address of 
the MCS queue element inserted by the owner thread is usually recorded in the lock 
instance so it can be conveyed to the subsequent unlock operation to identify the successor, if any.
That field is protected by the lock itself and accessed within the critical section.  
Accesses to the field that records the owner's queue element address may themselves generate 
additional coherence traffic, although some implementations may avoid such accesses to 
shared fields by storing the queue element address in a thread-local associative structure that maps 
lock addresses to the owner's queue element address.

\Invisible{The standard POSIX pthread mutex APIs have no provision to pass data from the acquire
operation to the corresponding to unlock so typical MCS or CNA implementations adhering 
to that API are forced to either store the address of the owner's queue element in the 
lock structure itself, to convey that address to the unlock operator -- 
generating additional coherence traffic -- or use supplementary  per-thread maps that 
associate held locks with the owning queue element.} 

\BoldSection{CNA: } Compact NUMA-Aware locks (CNA) \cite{EuroSys19-CNA} are 
based on MCS, but add NUMA-awareness.   At arrival time, threads annotate 
their queue element with their NUMA node number.  At unlock-time, the owner 
scans forward into the primary MCS chain and culls remote elements,
transferring them to a secondary chain of remote threads.  
That secondary chain is propagated from the unlock operator to the successor via the 
queue elements, so the lock structure remains compact.  
Reducing the NUMA diversity of the primary chain acts to reduce 
\emph{lock migration}\cite{topc15-dice} and improve performance.  
To avoid indefinite starvation of threads on the secondary chain, the unlock 
operator periodically flushes the secondary chain back into the primary chain 
to shift  the currently preferred NUMA node.   At unlock-time, if the primary
chain is found empty, the secondary is flushed back into the primary to reprovision
the primary chain.  CNA unlock prefers to dispatch to threads on the primary, 
but will revert to the secondary list if the primary is empty.
The secondary chain is manipulated under the lock itself, in the
unlock operation.  
While CNA is NUMA-aware, compared to MCS, a number of additional CNA-specific
administrative steps -- culling, reprovisioning, periodic flushing -- execute 
while the lock is held and are subsumed into the critical section, 
potentially increasing the effective hold time of the lock.  
We observe that all NUMA-aware locks trade-off short-term fairness for improved
overall throughput. 


\section{The Fissile Algorithm} 

Fissile augments CNA with a TS fast-path using the \emph{LOITER} lock construction 
(\underline{L}ocking : \underline{O}uter-\underline{I}nner \underline{T}ranformation) 
\cite{arxiv-Malthusian} where the \emph{outer lock} is a TS lock and 
the \emph{inner lock} is a CNA lock.  
Acquiring ownership of the outer TS lock confers ownership of the compound Fissile lock.
Arriving threads first try the fast-path TS lock and, if successful, 
immediately enter the critical section. 
Otherwise control diverts into the slow path where the thread acquires the inner 
CNA lock.  We refer to the owner of the inner CNA lock as the \emph{alpha} thread.  
Once the CNA lock has been acquired, the alpha thread then busy-waits on the TS outer 
lock.  At most one thread at any one time busy-waits on the outer TS lock,
avoiding the scalability impact of global spinning,
where multiple threads simultaneously busy-wait on a given location.  
As there is at most one thread busy-waiting on the outer lock, we use TS instead of TTS.  
Once the outer lock has been acquired, we release the inner lock and enter the 
critical section.   To release a Fissile lock, we simply release the outer TS lock,
regardless of whether the corresponding acquisition took the fast path or slow path.  

\Invisible{Alpha; Alfa; OnDeck; first-among-equals; head-of-line; ready; standby;
Primus Inter Pares; Primo} 

\Invisible{Handover; response time; latency; convey; transfer; pass} 
\Invisible{Proactive; early; aggressive; prompt;} 
\Invisible{Decouple lock acquisition and queueing/waiting;} 
\Invisible{The owner of the outer TS lock is the owner of the compound Fissile lock.} 

A thread holds the inner CNA lock only within the Fissile lock acquisition operator.  
Specifically, Fissile releases the inner CNA lock within the Fissile acquire operation,
but while still holding the outer TS lock, potentially extending the hold-time of the outer lock.  
This choice, however, allows us to allocate the MCS queue element on-stack, which is a distinct advantage, 
avoiding MCS queue element allocation and deallocation. (Classic MCS requires one allocated queue element for 
each lock concurrently held by a thread whereas our approach avoids that expense). 
Furthermore the queue element of the alpha thread does not need to be communicated from the 
Fissile acquire operation to the unlock operation, as is the case for normal MCS and CNA.  
We employ a specialized CNA implementation, described below, which shifts much of the 
administrative overhead specific to CNA and normally found in the unlock operator to run 
before we acquire the outer TS lock, so the overhead of releasing the CNA inner lock 
while holding the outer TS lock is minimized.   

In Listing-\ref{Listing:Fissile-py} we provide a sketch of the Fissile algorithm.  
The \texttt{Outer} field is a TS lock word which can take on 3 values:
$0$ indicates \emph{unlocked}; $1$ indicates \emph{locked} and $2$ encodes
a special locked state used when the alpha thread is impatient and the previous
owner is transferring ownership of the outer TS lock directly to the alpha thread. 
\texttt{Inner} is the CNA inner lock, and \texttt{Impatient} reflects
the state of the alpha thread. 

\Invisible{preferred embodiment} 

\Invisible{
tension and trade-off: 
@  balance quality of admission schedule (NUMA) vs reorganization latency
@  fairness vs throughput : latitude and laxity; unfairness may be faster. 
}

Absent remediation, simple TS allows indefinite bypass and starvation of waiting threads. 
To avoid this issue, the alpha threads busy-waits on the TS lock for a short \emph{grace period}
but will then become ``impatient'' and cue direct handover of ownership the next time the TS lock 
is released, bounding bypass.  

When the alpha thread becomes impatient, having failed to acquire the outer lock
within the grace period, it sets the \texttt{Impatient}
field from the normal state of 0 to 2.  The unlock operator fetches from \texttt{Impatient} and stores
that value into the TS lock word.  
In typical circumstances when unlock runs after the
alpha has become impatient, it will observe and fetch 2 from \texttt{Impatient} and store that 
value into the TS lock word.  The alpha will then notice that the value 2 has propagate
from \texttt{Impatient} into the lock word, and takes direct handoff of ownership from 
that previous owner, restoring the lock word from 2 back to 1.  
If the unlock operation happens to run concurrently with the alpha thread becoming 
impatient, the unlock may race and fetch 0 from \texttt{Impatient} instead of 2.  
In this case either the alpha manages to seize the TS lock and acquire it when it 
becomes 0, or some other thread managed to pounce on the TS lock, in which case the
alpha thread must wait one more lock cycle to take ownership.  At worst, impatient
handover is delayed by one acquire-release cycle.  Once the value of 2 is visible 
to threads in unlock, immediate handover to the alpha is assured.  
Threads arriving in the fast-path that observe 2 will divert immediately into 
the CNA slow-path.  

The grace period serves as tunable parameter reflecting 
the trade-off and tension between throughput and short-term fairness. 
A shorter grace periods yields less bypass and fairer admission, while longer periods
may allow better throughput but worse short-term fairness.  

Fissile provides hybrid succession, employing \emph{competitive succession} \cite{arxiv-Malthusian}
when there is no contention, but switching to more conservative \emph{direct succession}
when the alpha thread becomes impatient.  Under competitive succession, the owner releases
the lock, allowing other waiting or arriving threads to acquire the lock.
Unfettered competitive succession admits undesirable long-term unfairness and starvation
but typically performs well under light load.  In addition, competitive succession
tends to provide more graceful throughput under preemption.  In direct succession,
as used by MCS, for instance, the lock holder directly transfers ownership to a waiting
successor without any intermediate or intervening transition to an unlocked state. 
All strict FIFO locks employ direct succession.  Direct succession suffers
under preemption, however, as ownership may be conveyed to a preempted thread,
and we have to wait for operating system time-slicing to dispatch the owner
onto a processor.

\Invisible{Barge; pounce; renouncement; Rapacious locks; } 
\Invisible{Alpha reverts to impatience} 
\Invisible{Hybrid; Composite; Compound}  
\Invisible{Proactively} 
\Invisible{Analogy : CSMA/CD - competitive succession ; Token Ring - direct succession} 

By restricting the number of threads competing for the outer TS lock, we improve
the odds that an arriving thread will find the lock clear and manage to acquire 
via the TS outer fast path.  Under fixed load, the system will tend to reach a balanced 
steady state where many circulating threads tend to acquire the TS lock without waiting.  

As shown in \cite{arxiv-TWA}, as more threads busy-wait on a given location, as
is the case in TS, stores to that location take longer to propagate.
(Concurrent reads to a given location scale, but concurrent writes or atomics do not \cite{pact15-schweizer}).  
Fissile addresses that concern by ensuring that only the alpha thread busy-waits
on the outer TS lock at any given time, accelerating handover.

The TS fast path provides the following benefits. 
First, latency is reduced, relative to MCS and CNA, for the uncontended case. 
Acquisition requires an atomic instruction and just a simple store to release. 
Second, the slow-path CNA MCS nodes can be allocated on-stack, simplifying 
the CNA implementation and avoiding the need to communicate or convey
the owner's MCS node from the lock operation to the corresponding unlock.
Third, TS with bounded bypass performs well under preemption, relative to MCS.  
Finally, and less obviously, the TS fast path provides benefit in the contended case.  
Fissile provides significant improvement over CNA when the critical section is small, 
and CNA has a hard time ``keeping up'' with the flow of arriving threads. 
That is, for very short critical sections, CNA itself -- CNA overheads --  
becomes the bottleneck for throughput \cite{isca10-eyerman}. 
Under intense contention the TS lock allows more throughput, 
serving as an alternative bypass channel, giving contention ``pressure'' 
a way to get around CNA when CNA becomes the bottleneck.   
When the critical sections are longer, fissile performs like CNA.   
Allowing some threads to pass through the CNA slow path and some fraction over
the TS fast path would appear to dilute CNA's NUMA benefits, but in 
practice, we find that CNA still quickly acts to filter out remote threads from
a set of threads circulating over a contended lock.  

\Invisible{Dilute; attenuate; } 

\Invisible{Fissile provides significant improvement over
CNA when the critical section is small, and CNA has a hard time ``keeping up''
with the flow of arriving threads. That is, for very short critical sections,
CNA itself -- CNA overheads --  becomes the bottleneck for throughput.
In this case the TS lock allows more throughput under intense contention,
as well as providing a low-latency fast-path under no or light contention.
Fissile TS serves bypass channel, giving the ``pressure'' a way to get around
CNA when CNA becomes the bottleneck.   When the critical sections are longer,
fissile performs like CNA.} 

The result is a highly scalable NUMA-aware lock that performs like TS at low 
contention and as well or better than CNA at high contention.    
Fissile provides short-term \emph{concurrency restriction}\cite{arxiv-Malthusian} which may 
improve overall throughput over a contended lock.  
Fissile locks are compact and also tolerate preemption, by virtue of the TS outer
lock, more gracefully than does CNA or MCS.  

\Invisible{In fact the performance of Fissile often exceeds that of either of its underlying lock types.} 
\Invisible{Synergy with max more than max of the parts} 

\Invisible{An atomic is required to acquire the lock and simple store to release.} 

\Invisible{Off-load contention into CNA; Divert; Deflect; relegate; } 

\subsection{Specialized CNA} 

\Invisible{Adapt CNA to fissile -- specialization} 

Classic CNA performs reorganization of the MCS chain -- to be more NUMA-friendly
and reduce NUMA lock transitions -- while
holding the the CNA lock itself, extending the effective critical section length and delaying
handover to a successor.  
Handover time impacts the scalability as the lock is held throughout handover, 
increasing the effective length of the critical section.
At extreme contention, the critical section length determines throughput
\cite{isca10-eyerman,podc18-aksenov}.
Fissile uses a specialized variant of CNA which reorganizes the chain of 
waiting threads early, immediately after acquiring the CNA lock.  
As such, reorganization runs outside and before the TS critical section, off the critical path, 
and potentially allows pipelining and overlap with the critical section execution.  
(Arguably, earlier reorganization may suffer as there are fewer threads enqueued from
which to schedule, but we have not observed any performance penalty related to this concern). 

The variant of CNA used by Fissile differs from the original \cite{EuroSys19-CNA} as follows.  
Classic CNA, at unlock-time, culls the entire remote suffix of the primary chain
into the remote list.  Our variant looks ahead only one thread into the primary MCS chain, 
and provides constant-time culling costs, yielding less potentially futile scanning
of the chain, and more predictable overheads.  In addition, our look-ahead-one policy
generates less coherence traffic accessing the MCS chain elements, as the element
examine for potential culling would also be accessed in the near future
when we subsequently release the CNA lock.

Finally, our version of CNA performs CNA administrative duties -- flushing and culling -- 
immediately after the owner acquires the CNA lock, whereas classic CNA defers those operations until
unlock-time.  Specifically, we reorganize outside and before the outer TS critical section,
allowing more overlap between CNA administrative duties and the execution of the critical section,
and accelerating CNA lock handover.  

All the changes above are optional optimizations and are not required to use CNA
within Fissile, but they serve to enhance performance.

\lstset{language=Python}
\lstset{frame=lines}
\lstset{caption={Simplified Pseudocode Implement of Fissile}}
\lstset{label={Listing:Fissile-py}}
\lstset{basicstyle=\fontfamily{fvm}\selectfont\footnotesize} 
\lstset{commentstyle=\itshape\color{gray}} 
\lstset{commentstyle=\slshape\color{gray}} 
\lstset{commentstyle=\itshape\color{gray}} 
\lstset{keywordstyle=\color{forestgreen}\bfseries} 
\lstset{backgroundcolor=\color{Gray95}} 


\lstset{language=Python,caption={Fissile Pseudocode},label={Listing:Fissile-py}}
\noindent\begin{figure}[htp]
\begin{lstlisting}[language=Python,frame=tb]
class Fissile :
  ## Outer : TTS Lock
  atomic<int> Outer     = 0 ; 

  ## anti-starvation
  atomic<int> Impatient = 0 ; 
  
  ## Inner : CNA-MCS lock  
  ## Represented by tail pointer
  CNAMCSLock Inner {} ; 

def Lock (Fissile * L) :
|  ## try fast-path outer : trylock
|  ## Thread that acquired the outer lock owns the
|  ## compound fissile lock
|  if AtomicCAS (&L->Outer, 0, 1) == 0 :
|  |  return 
|  
|  ## Contention : Revert to slow path ...
|  ## Acquire the inner CNA-MCS lock
|  ## MCS queue element allocated on-stack 
|  auto QueueElement I {} ; 
|
|  CNAAcquire (&L->Inner, &I) ; 
|  ## This thread holds the inner CNA lock
|  ## and assumes the role of the "alpha" thread.  
|
|  ## Execute CNA administrative operation to cull 
|  ## remotes from primary into remote or to flush
|  ## elements from remote into primary
|  CNACullOrFlush (&L->Inner, &I) ; 
|  
|  ## Acquire the outer lock
|  ## Patient waiting phase - grace period
|  ## allow bypass over outer TTS lock
|  for :
|  |  if AtomicSwap (&L->Outer, 1) == 0 : 
|  |  |  goto Exeunt
|  |  if PatienceExhausted() : break
|  |  Pause() ; 
|  
|  ## Enter impatient waiting phase 
|  ## suppress bypass to avoid starvation
|  assert L->Impatient == 0 ;
|  L->Impatient = 2 ;
|  for
|  |  ## Wait for swap to overwrite either 0 or 2 
|  |  if AtomicSwap (&L->Inner, 1) != 1 : 
|  |  |  break ; 
|  |  Pause() ; 
|  L->Impatient = 0 ; 
|  
|  Exeunt: 
|  ## We hold the outer lock
|  ## Release the inner lock
|  assert L->Outer == 1 ;
|  CNARelease (&L->Inner, &I) ; 
  
def Unlock (Fissile * L) :
|  ## Release the outer lock
|  assert L->Outer != 0 ;
|  L->Outer = L->Impatient ;

\end{lstlisting}
\end{figure}


\section{Related Work}

While mutual exclusion remains an active research topic 
\cite{podc18-aksenov,isca10-eyerman,ols12,sosp13-david,Middleware16-Antic,atc16-Guiroux,EuroPar19-GCR,usenixatc17-kashyap,sosp19-kashyap,tocs19-Guerraoui,cacm15-Bueso,icdcn20-jayanti} 
we focus on locks closely related to our design.

NUMA-aware locks attempt to restrict ownership of a lock
to threads on a given NUMA node over the short term, reducing so-called \emph{lock migration},
which can result in expensive inter-node coherence traffic.   
The first NUMA-aware lock was HBO (Hierarchical Back Off) \cite{HBO}, a test-and-set lock
where busy-waiting threads running on the same NUMA node as the current owner would use shorter
back-off durations, favoring the odds of handover to such proximal threads relative
to most distant threads.  While simple, HBO suffers from the same issues as other TS locks. 

Luchangco et al. \cite{europar06-luchangco} introducing HCLH, 
a NUMA-aware hierarchical version of the CLH queue-lock \cite{ipss94-magnusson,craig-clh}. 
The HCLH algorithm collects requests on each node into a local CLH-style queue, and 
then has the thread at the head of the queue integrate each node’s queue into a single global queue.
This avoids the overhead of spinning on a shared location and eliminates fairness and
starvation issues. HCLH intentionally inserts non work-conserving combining delays
to increase the size of groups of threads to be merged into the global queue. 
It was subsequently discovered that HCLH required threads to be bound to one 
processor for the thread’s lifetime.  Failure to bind could result in 
exclusion and progress failures, and as such we will not consider HCLH further. 

NUMA-aware Cohort locks \cite{PPoPP12-dice,topc15-dice} spawned various derivatives 
\cite{PPoPP15-Chabbi, PPoPP16-Chabbi}. 
While cohort locks scale well, they have a large variable-sized footprint.  
The size of a cohort lock instance is a function of the number of NUMA nodes, and is thus
not generally known until run-time, complicating static allocation of cohort locks. 
Being hierarchical in nature, they suffer increased latency under low or no contention 
as acquisition requires acquiring both node-level locks and top-level lock.  
CNA avoids all these concerns and is superior to cohort locks.
A changeset to convert the Linux kernel \emph{qspinlock} low-level spin lock
\cite{linux-locks,Long13} implementation from
an MCS-based design to CNA is under under submission at the time of writing
\footnote{\url{https://lwn.net/Articles/805655/}}.  Similarly, Fissile locks
are readily portable into the kernel environment.  

Kashyap's et al. \cite{sosp19-kashyap} \emph{Shuffling Lock} also performs NUMA-aware 
reorganization of MCS chains of waiters off the critical path, by waiting threads.  
They also use a LOITER-based design, but do not allow bypass. 
In the evaluation section, below, we compare Fissile against their user-mode 
implementation.


LOITER-base designs \cite{arxiv-Malthusian} first appeared, to our knowledge, in 
the HotSpot Java Virtual Machine implementation
\footnote{\url{https://github.com/openjdk-mirror/jdk7u-hotspot/blob/master/src/share/vm/runtime/mutex.cpp\#L168}}
in 2007.  The ``Go'' language runtime mutex\cite{GoMutex} uses a LOITER-based scheme
where the inner lock is implemented via a semaphore and time-bounded bypass is allowed. 
The linux kernel \emph{QSpinlock}\cite{Long13} construct also has a dual path TS and MCS lock, 
but does not allow bypass. The QSpinLock TS fast-path avoids MCS latency overheads in the uncontended case.  

Various authors \cite{ispan05-ha,Middleware16-Antic} have suggested switching adaptively
between MCS and lower latency locks depending on the contention level.  While workable, this adds
considerable algorithmic complexity, particularly for the changeover phase, and requires tuning.
\citet{asplos94-lim} suggested a more general framework for switching locks at runtime.

\Invisible{Concerns: reactivity response time; hysteresis and damping; chase; hunt; ring}
\Invisible{Parameter Parsimony} 


\section{Empirical Evaluation}
\label{section:Empirical} 

Unless otherwise noted, all data was collected on an Oracle X5-2 system.  
The system has 2 sockets, each populated with 
an Intel Xeon E5-2699 v3 CPU running at 2.30GHz.  Each socket has 18 cores, and each core is 2-way 
hyperthreaded, yielding 72 logical CPUs in total.  The system was running Ubuntu 18.04 with a stock 
Linux version 4.15 kernel, and all software was compiled using the provided GCC version 7.3 toolchain
at optimization level ``-O3''.  
64-bit C or C++ code was used for all experiments.  
Factory-provided system defaults were used in all cases, and Turbo mode \cite{turbo} was left enabled.  
In all cases default free-range unbound threads were used.  

We implemented all user-mode locks within LD\_PRELOAD interposition
libraries that expose the standard POSIX \texttt{pthread\_\allowbreak{}mutex\_t} programming interface
using the framework from ~\cite{topc15-dice}.  
This allows us to change lock implementations by varying the LD\_PRELOAD environment variable and
without modifying the application code that uses locks.
The C++ \texttt{std::mutex} construct maps directly to \texttt{pthread\_mutex} primitives,
so interposition works for both C and C++ code.
All busy-wait loops used the Intel \texttt{PAUSE} instruction.
We note that user-mode locks are not typically implemented as pure spin locks, instead
often using a spin-then-park waiting policy which voluntarily surrenders the CPUs of waiting threads
after a brief optimistic spinning period designed to reduce the context switching rate.
In our case, we find that user-mode is convenient venue for experiments, and note
in passing that threads in the CNA slow-path are easily made to park.

\Invisible{PAUSE for polite waiting} 

\Invisible{We use a 128 byte sector size on Intel processors for alignment to avoid 
false sharing.  The unit of coherence is 64 bytes throughout the cache hierarchy, but 128 bytes
is required because of the adjacent cache line prefetch facility where pairs of lines are automatically 
fetched together.  } 

\subsection{MutexBench}

\begin{figure}[h!]
\includegraphics[width=8.5cm]{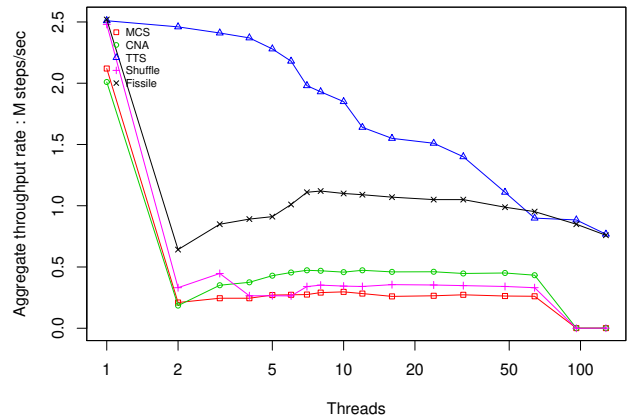}
\vspace{-18pt}      
\caption{MutexBench: Maximum Contention}
\label{Figure:MaximumContention}
\end{figure}

The MutexBench benchmark spawns $T$ concurrent threads. Each thread loops as follows: 
acquire a central lock L; execute a critical section; release L; execute
a non-critical section. At the end of a 10 second measurement interval the benchmark 
reports the total number of aggregate iterations completed by all the threads. 
We report the median of 7 independent runs in Figure-\ref{Figure:MaximumContention}.  
The critical section advances a C++ \texttt{std::mt19937} pseudo-random generator (PRNG) 2 steps.  
The non-critical section is empty.  
For clarity and to convey the maximum amount of information to allow a comparision the algorithms, 
the $X$-axis is offset to the minimum score and the $Y$-axis is logarithmic.
\Invisible{To facilitate comparison; visual comparison;} 

Immediately before acquiring the lock, each thread fetches the value of a shared
\emph{lock clock} value.  The critical section advances that value.  Subtracting
the clock value fetched in the critical section from the value fetched before acquiring
the lock gives a useful approximation of the thread's waiting time, given in
units of lock acquisitions.  Within the critical section, we record that waiting time
value into a global log.  After the measurement interval the benchmark harness post-processes 
the log to produce statistics describing the distribution of the waiting time values,
which reflect short-term fairness of the lock algorithm.  
The critical section also tallies lock migrations. 
These activities increase the effective length of the critical section.  

We ran the benchmark under the following lock algorithms: \textbf{TTS} is a simple
test-and-test-and-set lock using classic truncated randomized binary exponential back-off
\cite{tpds90-Anderson,tocs91-MellorCrummey} with the back-off duration capped to 100000 
iterations of a PAUSE loop;
\textbf{MCS} is classic MCS; \textbf{CNA} is described in \cite{EuroSys19-CNA} with
the probability of flushing the secondary chain into the primary configured as $P=1/256$
\footnote{We picked $P=1/256$ to match the default value used by the Shuffle Lock, allowing
a fair comparison between that lock and CNA};
\textbf{Shuffle} is Kashyap's \emph{Shuffle Lock}\cite{sosp19-kashyap} \emph{aqswonode} variant
\footnote{Taken verbatim from \url{https://github.com/sslab-gatech/shfllock/blob/master/ulocks/src/litl/src/aqswonode.c} and integrated into our LD\_PRELOAD framework}; 
\textbf{Fissile} is the Fissile algorithm described above with the grace period configured
as 50 steps of the TS loop executed by the alpha thread and the CNA flush probability
configured for $P=1/256$.  

\Invisible{CNA flush interval set to 50000 iterations of the waiting loop, with 
PAUSE instruction, executed by the head of the CNA secondary chain.} 

In Figure-\ref{Figure:MaximumContention} we make the following observations regarding
operation at maximal contention with an empty critical section:
\footnote{We note in passing that care must be taken when \emph{negative} or \emph{retrograde}
scaling occurs and aggregate performance degrades as we increase threads. 
As a throught experiment, if a hypothetical lock implementation were to introduce 
additional synthetic delays outside the critical path, aggregate performance might increase as 
the delay throttles the arrival rate and concurrency over the contended lock.  
As such, evaluating just the maximal contention case in isolation is insufficient.}. 
\begin{itemize}[align=left,leftmargin=2em,labelwidth=0.8em]
\item At 1 thread the benchmark measures the latency of uncontended acquire and release operations.  
MCS and CNA lag behind TTS, Shuffle and Fissile as they lack a fast-path. 
\item At or above 2 threads, most algorithms fall behind TTS as TTS starves all but one thread
for long periods, effectively yielding performance near that found at just one thread.
\item Broadly, Fissile outperforms CNA and CNA outperforms Shuffle.
\item Above 72 threads we encounter preemption via time slicing.  TTS and Fissile are
tolerant of preemption where the other forms with direct handover encounter
a precipitous drop in performance.  
\end{itemize} 


\Invisible{Drill-down; exemplar; T=10 } 

In Table-\ref{Table:Detailed} we provide additional details for execution at 10 threads.
\textbf{Throughput} is given in units of millions of acquires per second aggregate
throughput for all threads; 
\textbf{Spread} reflects long-term fairness between threads, computed as the maximum 
number of iterations completed by any thread within the measurement interval divided by the minimum; 
\textbf{Migration} is the reciprocal of the NUMA lock migration rate.  
(A Migration value of $N$ indicates that the lock migrated between NUMA nodes 1 out of
every $N$ lock acquisitions, on average).  
The remaining columns describe the distribution of the observed waiting times,
which we use to measure short-term fairness.
\textbf{RSTDDEV} is the relative standard deviation \cite{rstddev}; 
\textbf{Theil-T} is the normalized Theil-T index \cite{Theil, Theil-w} -- used 
in the field of econometrics as a metric of income disparity and unfairness -- where
a value of 0 is ideally fair and 1 is maximally unfair.  

\Invisible{Disperal; Disparity; Diverity; inequality; uniformity; unfairness;} 

We observe that TTS is deeply unfair over the long term and short term. 
TTS also exhibits a surprisingly low lock migration rate -- on average 1 migration per 323 acquisitions -- 
presumably arising
from platform-specific cache line arbitration phenomena.  Somewhat perversely, this makes
TTS implicitly NUMA-friendly, reducing migration rates.  
TTS is vulnerable to the \emph{Matthew Effect}\cite{MatthewEffect}\footnote{Sometimes called the \emph{capture effect}} --
once a thread has entered deeper back-off, it is less likely to acquire the 
lock in unit time, amplifying subsequent unfairness.  
The remaining locks show reasonable long-term and short-term fairness.

\begin{figure}[h!]
\includegraphics[width=8.5cm]{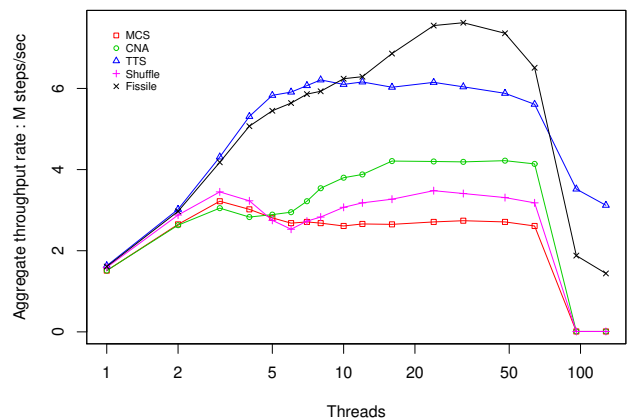}
\vspace{-18pt}      
\caption{MutexBench: Moderate Contention}
\label{Figure:ModerateContention}
\end{figure}

In Figure-\ref{Figure:ModerateContention} we configure the benchmark so the non-critical section 
generates a uniformly distributed random value in $[0-200)$ and steps the
thread-local random number generator that many steps, admitting potential
positive scalability.   In this moderate contention case we can see
that Fissile and TTS locks tend to provide the best performance, although
the TTS lock is again unfair.  
Shuffle, CNA, and Fissile show a positive inflection around 12 threads, as there are
sufficient waiting threads to allow NUMA-friendly intra-node handover. 
Again, we see an abrupt drop in throughput above 72 threads when preemption is active,
but note that Fissile and TTS more gracefully tolerate preemption. 

\Invisible{Methodological trap and flaw when negative scaling occurs; 
Tuning the lock in isolution in this mode -- at maximal contention with
an empty non-critical section -- can lead us to incorrectly tune toward 
making the lock ``worse'' as ensuing inefficient delays may reduce the arrival rate
and throttle concurrency, thus making aggregate performance better
in the maximal contention case as we worsen the lock for the general case.
Deceptive and confounding; peril; misleading; 
\url{https://blogs.oracle.com/dave/the-perils-of-negative-scalability}
\url{https://en.wikipedia.org/wiki/Brooks's_law}
\url{https://arxiv.org/abs/1506.06796} SIF = Slower-is-faster
} 

\newcolumntype{a}{>{\columncolor{Gray95}}l}

\begin{table} [h]
\centering
{\fontsize{6.5}{6.5}\selectfont
\begin{tabular}{lccccc}
\toprule

&
\multicolumn{1}{c}{Throughput}      &
\multicolumn{1}{c}{Spread}          &
\multicolumn{1}{c}{Migration}       &
\multicolumn{1}{c}{RSTDDEV}         &
\multicolumn{1}{c}{Theil-T}         \\
\midrule

MCS       & .297  & 1.00    & 1.83 &  0.01    & 0.00   \\ 
CNA       & .458  & 1.06    & 254  &  13.5    & 0.17   \\
TTS       & 1.85  & 7.89    & 323  &  102     & 0.44   \\
Shuffle   & .344  & 1.86    & 234  &  11.3    & 0.15   \\
Fissile   & 1.11  & 1.26    & 374  &  11.8    & 0.17   \\

\midrule[\heavyrulewidth]
\bottomrule
\end{tabular}%
}
\caption{Detailed Execution Analysis}\label{Table:Detailed}
\end{table}

\subsection{std::atomic} 

\begin{figure}[h!]
\includegraphics[width=8.5cm]{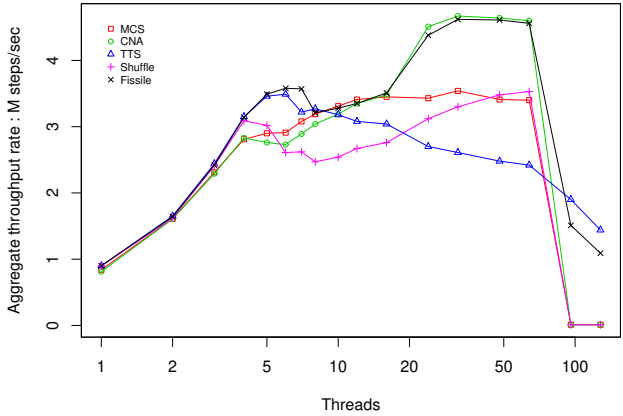}
\vspace{-18pt}      
\caption{C++ std::atomic}
\label{Figure:atomic}
\end{figure}

In Figure-\ref{Figure:atomic} we use a benchmark harness similar to that of
MutexBench but with the following differences.  The non-critical section uses a 
thread-local \texttt{std::mt19937} pseudo-random number generator (PRNG) to compute a value 
distributed uniformly in $[0,200)$ and then advances the PRNG that many steps.  
Instead of an explicit critical section, each iteration executes 
\texttt{A.load()} where A is shared an instance of \texttt{std::atomic<T>} and \texttt{T} 
is a simple \emph{struct} containing 5 32-bit integer fields.
The C++ compiler and runtime implement \texttt{std::atomic} for such objects by hashing
the address of the instance into an array of mutexes, and acquiring those as needed to implement
the desired atomic action.  Interestingly, the NUMA-aware locks, CNA, Shuffle and Fissile,
exhibit fading performance between 5 and 10 threads, but performance recovers at higher thread 
counts when there are sufficient waiting threads to profitably reorder for a NUMA-friendly admission
schedule. 
Below 10 threads, contention is sufficiently low that Fissile exceeds CNA by virtue of its fast-path.
Fissile and TTS provide similar performance in this region.  
Above 10 threads, the critical section is sufficiently long in duration that CNA and Fissile yield 
approximately the same performance.  

\begin{figure}[h!]
\includegraphics[width=8.5cm]{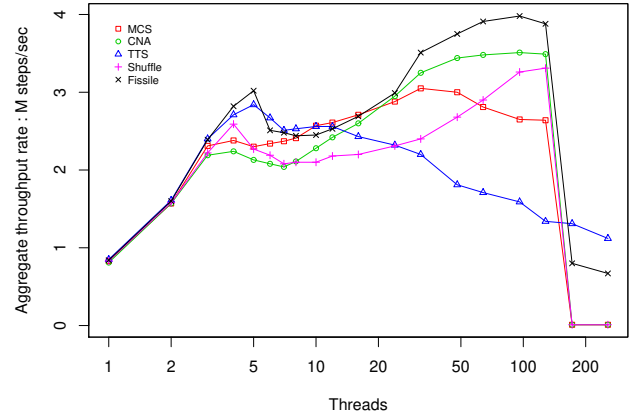}
\vspace{-18pt}      
\caption{C++ std::atomic on 4-node System}
\label{Figure:atomic-x54}
\end{figure}

In Figure-\ref{Figure:atomic-x54} we repeat the experiment in Figure-\ref{Figure:atomic}
on an Oracle X5-4, which has 4 NUMA nodes, 18 cores per socket and 2 hyperthreads
per core, for 144 logical CPUs, demonstrating that our approach generalizes to larger
NUMA systems.   The onset of benefit provided by NUMA-aware locks is somewhat delayed
as we have 4 nodes instead of 2 and, at a given thread count, threads are more dispersed
and the socket is less populated.  

\subsection{FIFO Support}

Fissile allows bypass both over the outer lock and within the CNA inner lock.
We can, however, easily modify Fissile to provide expedited FIFO-like admission service as follows.  
First, FIFO locking requests that pass into the slow path mark their CNA MCS 
queue element with a ``FIFO'' flag.  
CNA culling refrains from shifting such elements into the CNA secondary list.  
Critically, if element $S$ is marked as FIFO, then no requests that arrive after 
$S$ on the inner CNA lock will acquire that lock before $S$. 
We also suppress bypass over the outer lock while FIFO requests are waiting.
To that end, instead of setting and clearing the \texttt{Impatient} field we 
modify Fissile slightly to atomically fetch-and-add \texttt{Impatient} by 2 or -2, respectively.
(We also make a corresponding change to the comparison in the grace period loop from \texttt{== 0} to \texttt{!= 1}). 
When a FIFO request diverts into the slow path, it increments \texttt{Impatient} by 2 
before acquiring the CNA inner lock, and decrements by 2 after acquiring the outer lock. 
The request will be serviced in FIFO order, without being bypassed by more recently arrived threads,
once it increments \texttt{Impatient} --
and that value has become visible to threads in the unlock path -- and has
executed the \texttt{SWAP} instruction that appends the request to the CNA MCS chain.

To avoid fairness anomalies and make fairness analysis more tractable, we explicitly
do no change the preferred NUMA when when servicing a FIFO request. 

To demonstrate the efficacy of FIFO-enabled Fissile, we extended the \texttt{MutexBench}
benchmark harness to allow a mixture of normal and FIFO-designated threads,
both competing for a common lock.  
We used 25 normal threads, and 2 FIFO threads.  Normal threads advance the
global PRNG 2 times in the critical section, as described above, and in the non-critical section
compute a uniformly distributed random number in $[0-100)$ and advance a thread-local
PRNG instance that many steps.  FIFO threads execute the same critical section,
but use a non-critical section duration  randomly selected from the range $[0-2000)$, reflecting
intermittent low duty-cycle FIFO operations.  
The FIFO attribute is per-thread (but could also be specified for individual locking operations) 
and is ignored by all lock implementations except FIFO-enabled Fissile.  
All FIFO data was taken on the X5-2.    

Table-\ref{Table:FIFO} shows the results, comparing Fissile, FIFO-enabled Fissile, and MCS.  
We report throughput over a 10 second measurement interval broken out for
the normal threads and the FIFO threads.  We also report statistics describing the 
observed wait times, computed in logical \emph{lock clock} units, for the FIFO threads in isolation. 
As we can see \texttt{Fissile+FIFO} provides wait times very close to that afforded
by MCS, and with greater throughput for both normal and FIFO threads. 

\begin{table} [h]
\centering
{\fontsize{7.5}{7.5}\selectfont
\begin{tabular}{lcccccc}
\toprule
\multicolumn{1}{l}{} &
\multicolumn{2}{c}{Throughput} &
\multicolumn{4}{c}{Wait times for FIFO}  \\
\cmidrule(lr){2-3}
\cmidrule(lr){4-7}

&
\multicolumn{1}{c}{FIFO}    &
\multicolumn{1}{c}{Normal}  &
\multicolumn{1}{c}{RSTDDEV} &
\multicolumn{1}{c}{Worst}   &
\multicolumn{1}{c}{Avg}     & 
\multicolumn{1}{c}{Median}  \\
\midrule

MCS            & 1.3M & 23.0M   & 0.03 & 29     & 24.4  & 25 \\
Fissile        & 1.5M & 43.9M   & 52.3 & 531294 & 40.7  & 15 \\
Fissile+FIFO   & 2.7M & 38.8M   & 0.33 & 41     & 11.9  & 12 \\

\midrule[\heavyrulewidth]
\bottomrule
\end{tabular}%
}
\caption{FIFO Performance}\label{Table:FIFO}
\end{table}

\renewenvironment{quotation}%
  {\list{}{\leftmargin=0.16in\rightmargin=0.16in}\item[]}%
  {\endlist}


\section{Conclusion}

Fissile locks are compact, NUMA-aware, preemption tolerant, and scalable, but also provide
excellent latency at low or no contention.  The algorithm is straightforward and easily
integrated into existing locking infrastructures.   They are particularly helpful
under contention with high arrival rates and short critical sections.  
Contended locking uses the CNA lock while uncontended operations use the TS lock.
Fissile locks deflect contention away from TS lock into the CNA lock.

Bypass over the outer lock via the fast path is the key to Fissile.  
While the slow path provides a higher quality NUMA-friendly admission schedule,
it also suffers higher latency arising from the more complex lock mechanism. 
The fast path allows for low latency in the uncontended case, 
but also improves scalability under contention by augmenting the slow path with
an alternative if the the slow path lock overheads prove a bottleneck.  

\Invisible{Imposing no surprising or onerous requirements} 

\Invisible{In the Appendix we identify a number of variations on the basic Fissile algorithm that we plan to
explore in the future.}

\AtFoot{An extended version of this paper is available at \url{https://arxiv.org/abs/2003.05025}} 



\bibliography{Fissile.bib}


\begin{thebibliography}{42}


\ifx \showCODEN    \undefined \def \showCODEN     #1{\unskip}     \fi
\ifx \showDOI      \undefined \def \showDOI       #1{#1}\fi
\ifx \showISBNx    \undefined \def \showISBNx     #1{\unskip}     \fi
\ifx \showISBNxiii \undefined \def \showISBNxiii  #1{\unskip}     \fi
\ifx \showISSN     \undefined \def \showISSN      #1{\unskip}     \fi
\ifx \showLCCN     \undefined \def \showLCCN      #1{\unskip}     \fi
\ifx \shownote     \undefined \def \shownote      #1{#1}          \fi
\ifx \showarticletitle \undefined \def \showarticletitle #1{#1}   \fi
\ifx \showURL      \undefined \def \showURL       {\relax}        \fi
\providecommand\bibfield[2]{#2}
\providecommand\bibinfo[2]{#2}
\providecommand\natexlab[1]{#1}
\providecommand\showeprint[2][]{arXiv:#2}

\bibitem[\protect\citeauthoryear{??}{GoM}{2020}]%
        {GoMutex}
 \bibinfo{year}{2020}\natexlab{}.
\newblock \bibinfo{title}{Go RunTime : mutex implementation}.
\newblock
\newblock
\urldef\tempurl%
\url{https://github.com/golang/go/blob/master/src/sync/mutex.go}
\showURL{%
\tempurl}


\bibitem[\protect\citeauthoryear{Aksenov, Alistarh, and Kuznetsov}{Aksenov
  et~al\mbox{.}}{2018}]%
        {podc18-aksenov}
\bibfield{author}{\bibinfo{person}{Vitaly Aksenov}, \bibinfo{person}{Dan
  Alistarh}, {and} \bibinfo{person}{Petr Kuznetsov}.}
  \bibinfo{year}{2018}\natexlab{}.
\newblock \showarticletitle{Brief Announcement: Performance Prediction for
  Coarse-Grained Locking}. In \bibinfo{booktitle}{\emph{Proceedings of the 2018
  ACM Symposium on Principles of Distributed Computing}}
  \emph{(\bibinfo{series}{PODC '18})}. \bibinfo{publisher}{ACM}.
\newblock
\urldef\tempurl%
\url{https://doi.org/10.1145/3212734.3212785}
\showDOI{\tempurl}


\bibitem[\protect\citeauthoryear{Anderson}{Anderson}{1990}]%
        {tpds90-Anderson}
\bibfield{author}{\bibinfo{person}{T.~E. Anderson}.}
  \bibinfo{year}{1990}\natexlab{}.
\newblock \showarticletitle{The performance of spin lock alternatives for
  shared-money multiprocessors}.
\newblock \bibinfo{journal}{\emph{IEEE Transactions on Parallel and Distributed
  Systems}} (\bibinfo{year}{1990}).
\newblock
\urldef\tempurl%
\url{https://doi.org/10.1109/71.80120}
\showDOI{\tempurl}


\bibitem[\protect\citeauthoryear{Anti\'{c}, Chatzopoulos, Guerraoui, and
  Trigonakis}{Anti\'{c} et~al\mbox{.}}{2016}]%
        {Middleware16-Antic}
\bibfield{author}{\bibinfo{person}{Jelena Anti\'{c}}, \bibinfo{person}{Georgios
  Chatzopoulos}, \bibinfo{person}{Rachid Guerraoui}, {and}
  \bibinfo{person}{Vasileios Trigonakis}.} \bibinfo{year}{2016}\natexlab{}.
\newblock \showarticletitle{Locking Made Easy}. In
  \bibinfo{booktitle}{\emph{Proceedings of the 17th International Middleware
  Conference}} \emph{(\bibinfo{series}{Middleware '16})}.
  \bibinfo{publisher}{ACM}.
\newblock
\urldef\tempurl%
\url{https://doi.org/10.1145/2988336.2988357}
\showDOI{\tempurl}


\bibitem[\protect\citeauthoryear{Boyd-Wickizer, Kaashoek, Morris, and
  Zeldovich}{Boyd-Wickizer et~al\mbox{.}}{2012}]%
        {ols12}
\bibfield{author}{\bibinfo{person}{Silas Boyd-Wickizer},
  \bibinfo{person}{M.~Frans Kaashoek}, \bibinfo{person}{Robert Morris}, {and}
  \bibinfo{person}{Nickolai Zeldovich}.} \bibinfo{year}{2012}\natexlab{}.
\newblock \showarticletitle{Non-scalable locks are dangerous}.
\newblock \bibinfo{journal}{\emph{Ottawa Linux Symposium (OLS)}}
  (\bibinfo{year}{2012}).
\newblock
\urldef\tempurl%
\url{https://www.kernel.org/doc/ols/2012/ols2012-zeldovich.pdf}
\showURL{%
\tempurl}


\bibitem[\protect\citeauthoryear{Bueso}{Bueso}{2014}]%
        {cacm15-Bueso}
\bibfield{author}{\bibinfo{person}{Davidlohr Bueso}.}
  \bibinfo{year}{2014}\natexlab{}.
\newblock \showarticletitle{Scalability Techniques for Practical
  Synchronization Primitives}.
\newblock \bibinfo{journal}{\emph{Commun. ACM}} (\bibinfo{year}{2014}).
\newblock
\urldef\tempurl%
\url{http://doi.acm.org/10.1145/2687882}
\showURL{%
\tempurl}


\bibitem[\protect\citeauthoryear{Chabbi, Fagan, and Mellor-Crummey}{Chabbi
  et~al\mbox{.}}{2015}]%
        {PPoPP15-Chabbi}
\bibfield{author}{\bibinfo{person}{Milind Chabbi}, \bibinfo{person}{Michael
  Fagan}, {and} \bibinfo{person}{John Mellor-Crummey}.}
  \bibinfo{year}{2015}\natexlab{}.
\newblock \showarticletitle{High Performance Locks for Multi-Level NUMA
  Systems}. In \bibinfo{booktitle}{\emph{Proceedings of the 20th ACM SIGPLAN
  Symposium on Principles and Practice of Parallel Programming}}
  \emph{(\bibinfo{series}{PPoPP 2015})}. \bibinfo{publisher}{Association for
  Computing Machinery}.
\newblock
\urldef\tempurl%
\url{https://doi.org/10.1145/2688500.2688503}
\showDOI{\tempurl}


\bibitem[\protect\citeauthoryear{Chabbi and Mellor-Crummey}{Chabbi and
  Mellor-Crummey}{2016}]%
        {PPoPP16-Chabbi}
\bibfield{author}{\bibinfo{person}{Milind Chabbi} {and} \bibinfo{person}{John
  Mellor-Crummey}.} \bibinfo{year}{2016}\natexlab{}.
\newblock \showarticletitle{Contention-Conscious, Locality-Preserving Locks}.
  In \bibinfo{booktitle}{\emph{Proceedings of the 21st ACM SIGPLAN Symposium on
  Principles and Practice of Parallel Programming}}
  \emph{(\bibinfo{series}{PPoPP ’16})}. \bibinfo{publisher}{Association for
  Computing Machinery}.
\newblock
\urldef\tempurl%
\url{https://doi.org/10.1145/2851141.2851166}
\showDOI{\tempurl}


\bibitem[\protect\citeauthoryear{Corbet}{Corbet}{[n. d.]a}]%
        {linux-page-struct}
\bibfield{author}{\bibinfo{person}{Jonathan Corbet}.} \bibinfo{year}{[n.
  d.]}\natexlab{a}.
\newblock \bibinfo{title}{Cramming more into struct page}.
\newblock \bibinfo{howpublished}{\url{https://lwn.net/Articles/565097}, August
  28, 2013}.
\newblock
\newblock
\shownote{Accessed: 2018-10-01.}


\bibitem[\protect\citeauthoryear{Corbet}{Corbet}{[n. d.]b}]%
        {linux-locks}
\bibfield{author}{\bibinfo{person}{Jonathan Corbet}.} \bibinfo{year}{[n.
  d.]}\natexlab{b}.
\newblock \bibinfo{title}{{MCS} locks and qspinlocks}.
\newblock \bibinfo{howpublished}{\url{https://lwn.net/Articles/590243}, March
  11, 2014}.
\newblock
\newblock
\shownote{Accessed: 2018-09-12.}


\bibitem[\protect\citeauthoryear{Corbet}{Corbet}{[n. d.]c}]%
        {ticket-spinlocks}
\bibfield{author}{\bibinfo{person}{Jonathan Corbet}.} \bibinfo{year}{[n.
  d.]}\natexlab{c}.
\newblock \bibinfo{title}{Ticket Spinlocks}.
\newblock \bibinfo{howpublished}{\url{https://lwn.net/Articles/267968},
  February 6, 2008}.
\newblock
\newblock
\shownote{Accessed: 2018-09-12.}


\bibitem[\protect\citeauthoryear{Craig}{Craig}{1993}]%
        {craig-clh}
\bibfield{author}{\bibinfo{person}{Travis Craig}.}
  \bibinfo{year}{1993}\natexlab{}.
\newblock \bibinfo{title}{Building FIFO and priority-queueing spin locks from
  atomic swap}.
\newblock
\newblock


\bibitem[\protect\citeauthoryear{David, Guerraoui, and Trigonakis}{David
  et~al\mbox{.}}{2013}]%
        {sosp13-david}
\bibfield{author}{\bibinfo{person}{Tudor David}, \bibinfo{person}{Rachid
  Guerraoui}, {and} \bibinfo{person}{Vasileios Trigonakis}.}
  \bibinfo{year}{2013}\natexlab{}.
\newblock \showarticletitle{Everything You Always Wanted to Know About
  Synchronization but Were Afraid to Ask}. In
  \bibinfo{booktitle}{\emph{Proceedings of the Twenty-Fourth ACM Symposium on
  Operating Systems Principles}} \emph{(\bibinfo{series}{SOSP '13})}.
\newblock
\urldef\tempurl%
\url{http://doi.acm.org/10.1145/2517349.2522714}
\showURL{%
\tempurl}


\bibitem[\protect\citeauthoryear{Dice}{Dice}{2015}]%
        {arxiv-Malthusian}
\bibfield{author}{\bibinfo{person}{Dave Dice}.}
  \bibinfo{year}{2015}\natexlab{}.
\newblock \showarticletitle{Malthusian Locks}.
\newblock \bibinfo{journal}{\emph{CoRR}}  \bibinfo{volume}{abs/1511.06035}
  (\bibinfo{year}{2015}).
\newblock
\showeprint[arxiv]{1511.06035}
\urldef\tempurl%
\url{http://arxiv.org/abs/1511.06035}
\showURL{%
\tempurl}


\bibitem[\protect\citeauthoryear{Dice}{Dice}{2017}]%
        {eurosys17-dice}
\bibfield{author}{\bibinfo{person}{Dave Dice}.}
  \bibinfo{year}{2017}\natexlab{}.
\newblock \showarticletitle{Malthusian Locks}. In
  \bibinfo{booktitle}{\emph{Proceedings of the Twelfth European Conference on
  Computer Systems}} \emph{(\bibinfo{series}{EuroSys '17})}.
\newblock
\urldef\tempurl%
\url{http://doi.acm.org/10.1145/3064176.3064203}
\showURL{%
\tempurl}


\bibitem[\protect\citeauthoryear{Dice and Kogan}{Dice and Kogan}{2018a}]%
        {arxiv-CNA}
\bibfield{author}{\bibinfo{person}{Dave Dice} {and} \bibinfo{person}{Alex
  Kogan}.} \bibinfo{year}{2018}\natexlab{a}.
\newblock \showarticletitle{Compact NUMA-Aware Locks}.
\newblock \bibinfo{journal}{\emph{CoRR}}  \bibinfo{volume}{abs/1810.05600}
  (\bibinfo{year}{2018}).
\newblock
\urldef\tempurl%
\url{http://arxiv.org/abs/1810.05600}
\showURL{%
\tempurl}


\bibitem[\protect\citeauthoryear{Dice and Kogan}{Dice and Kogan}{2018b}]%
        {arxiv-TWA}
\bibfield{author}{\bibinfo{person}{Dave Dice} {and} \bibinfo{person}{Alex
  Kogan}.} \bibinfo{year}{2018}\natexlab{b}.
\newblock \showarticletitle{{TWA} - Ticket Locks Augmented with a Waiting
  Array}.
\newblock \bibinfo{journal}{\emph{CoRR}}  \bibinfo{volume}{abs/1810.01573}
  (\bibinfo{year}{2018}).
\newblock
\showeprint[arxiv]{1810.01573}
\urldef\tempurl%
\url{http://arxiv.org/abs/1810.01573}
\showURL{%
\tempurl}


\bibitem[\protect\citeauthoryear{Dice and Kogan}{Dice and Kogan}{2019a}]%
        {EuroPar19-GCR}
\bibfield{author}{\bibinfo{person}{Dave Dice} {and} \bibinfo{person}{Alex
  Kogan}.} \bibinfo{year}{2019}\natexlab{a}.
\newblock \showarticletitle{Avoiding Scalability Collapse by Restricting
  Concurrency}. In \bibinfo{booktitle}{\emph{Euro-Par 2019: Parallel Processing
  - 25th International Conference on Parallel and Distributed Computing,
  G{\"{o}}ttingen, Germany, August 26-30, 2019, Proceedings}}
  \emph{(\bibinfo{series}{Lecture Notes in Computer Science})}.
  \bibinfo{publisher}{Springer}.
\newblock
\urldef\tempurl%
\url{https://doi.org/10.1007/978-3-030-29400-7\_26}
\showDOI{\tempurl}


\bibitem[\protect\citeauthoryear{Dice and Kogan}{Dice and Kogan}{2019b}]%
        {EuroSys19-CNA}
\bibfield{author}{\bibinfo{person}{Dave Dice} {and} \bibinfo{person}{Alex
  Kogan}.} \bibinfo{year}{2019}\natexlab{b}.
\newblock \showarticletitle{Compact NUMA-Aware Locks}. In
  \bibinfo{booktitle}{\emph{Proceedings of the Fourteenth EuroSys Conference
  2019}} \emph{(\bibinfo{series}{EuroSys ’19})}.
  \bibinfo{publisher}{Association for Computing Machinery}.
\newblock
\urldef\tempurl%
\url{https://doi.org/10.1145/3302424.3303984}
\showDOI{\tempurl}


\bibitem[\protect\citeauthoryear{Dice, Marathe, and Shavit}{Dice
  et~al\mbox{.}}{2012}]%
        {PPoPP12-dice}
\bibfield{author}{\bibinfo{person}{David Dice}, \bibinfo{person}{Virendra~J.
  Marathe}, {and} \bibinfo{person}{Nir Shavit}.}
  \bibinfo{year}{2012}\natexlab{}.
\newblock \showarticletitle{Lock Cohorting: A General Technique for Designing
  NUMA Locks}. In \bibinfo{booktitle}{\emph{Proceedings of the 17th ACM SIGPLAN
  Symposium on Principles and Practice of Parallel Programming}}
  \emph{(\bibinfo{series}{PPoPP ’12})}. \bibinfo{publisher}{Association for
  Computing Machinery}.
\newblock
\urldef\tempurl%
\url{https://doi.org/10.1145/2145816.2145848}
\showDOI{\tempurl}


\bibitem[\protect\citeauthoryear{Dice, Marathe, and Shavit}{Dice
  et~al\mbox{.}}{2015}]%
        {topc15-dice}
\bibfield{author}{\bibinfo{person}{David Dice}, \bibinfo{person}{Virendra~J.
  Marathe}, {and} \bibinfo{person}{Nir Shavit}.}
  \bibinfo{year}{2015}\natexlab{}.
\newblock \showarticletitle{Lock Cohorting: A General Technique for Designing
  NUMA Locks}.
\newblock \bibinfo{journal}{\emph{ACM Trans. Parallel Comput.}}
  (\bibinfo{year}{2015}).
\newblock
\urldef\tempurl%
\url{https://doi.org/10.1145/2686884}
\showDOI{\tempurl}


\bibitem[\protect\citeauthoryear{Eyerman and Eeckhout}{Eyerman and
  Eeckhout}{2010}]%
        {isca10-eyerman}
\bibfield{author}{\bibinfo{person}{Stijn Eyerman} {and} \bibinfo{person}{Lieven
  Eeckhout}.} \bibinfo{year}{2010}\natexlab{}.
\newblock \showarticletitle{Modeling Critical Sections in Amdahl's Law and Its
  Implications for Multicore Design}. In \bibinfo{booktitle}{\emph{Proceedings
  of the 37th Annual International Symposium on Computer Architecture}}
  \emph{(\bibinfo{series}{ISCA '10})}. \bibinfo{publisher}{ACM}.
\newblock
\urldef\tempurl%
\url{https://doi.org/10.1145/1815961.1816011}
\showDOI{\tempurl}


\bibitem[\protect\citeauthoryear{Guerraoui, Guiroux, Lachaize, Qu\'{e}ma, and
  Trigonakis}{Guerraoui et~al\mbox{.}}{2019}]%
        {tocs19-Guerraoui}
\bibfield{author}{\bibinfo{person}{Rachid Guerraoui}, \bibinfo{person}{Hugo
  Guiroux}, \bibinfo{person}{Renaud Lachaize}, \bibinfo{person}{Vivien
  Qu\'{e}ma}, {and} \bibinfo{person}{Vasileios Trigonakis}.}
  \bibinfo{year}{2019}\natexlab{}.
\newblock \showarticletitle{Lock–Unlock: Is That All? A Pragmatic Analysis of
  Locking in Software Systems}.
\newblock \bibinfo{journal}{\emph{ACM Trans. Comput. Syst.}}
  (\bibinfo{year}{2019}).
\newblock
\urldef\tempurl%
\url{https://doi.org/10.1145/3301501}
\showDOI{\tempurl}


\bibitem[\protect\citeauthoryear{Guiroux, Lachaize, and Qu\'{e}ma}{Guiroux
  et~al\mbox{.}}{2016}]%
        {atc16-Guiroux}
\bibfield{author}{\bibinfo{person}{Hugo Guiroux}, \bibinfo{person}{Renaud
  Lachaize}, {and} \bibinfo{person}{Vivien Qu\'{e}ma}.}
  \bibinfo{year}{2016}\natexlab{}.
\newblock \showarticletitle{Multicore Locks: The Case Is Not Closed Yet}. In
  \bibinfo{booktitle}{\emph{2016 {USENIX} Annual Technical Conference ({USENIX}
  {ATC} 16)}}. \bibinfo{publisher}{{USENIX} Association}.
\newblock
\showISBNx{978-1-931971-30-0}
\urldef\tempurl%
\url{https://www.usenix.org/conference/atc16/technical-sessions/presentation/guiroux}
\showURL{%
\tempurl}


\bibitem[\protect\citeauthoryear{Ha, Papatriantafilou, and Tsigas}{Ha
  et~al\mbox{.}}{2005}]%
        {ispan05-ha}
\bibfield{author}{\bibinfo{person}{P.~H. Ha}, \bibinfo{person}{M.
  Papatriantafilou}, {and} \bibinfo{person}{P. Tsigas}.}
  \bibinfo{year}{2005}\natexlab{}.
\newblock \showarticletitle{Reactive spin-locks: a self-tuning approach}. In
  \bibinfo{booktitle}{\emph{8th International Symposium on Parallel
  Architectures,Algorithms and Networks (ISPAN'05)}}.
\newblock
\urldef\tempurl%
\url{https://doi.org/10.1109/ISPAN.2005.73}
\showDOI{\tempurl}


\bibitem[\protect\citeauthoryear{Jayanti, Jayanti, and Jayanti}{Jayanti
  et~al\mbox{.}}{2020}]%
        {icdcn20-jayanti}
\bibfield{author}{\bibinfo{person}{Prasad Jayanti}, \bibinfo{person}{Siddhartha
  Jayanti}, {and} \bibinfo{person}{Sucharita Jayanti}.}
  \bibinfo{year}{2020}\natexlab{}.
\newblock \showarticletitle{Towards an Ideal Queue Lock}. In
  \bibinfo{booktitle}{\emph{Proceedings of the 21st International Conference on
  Distributed Computing and Networking}} \emph{(\bibinfo{series}{ICDCN 2020})}.
  \bibinfo{publisher}{Association for Computing Machinery}.
\newblock
\urldef\tempurl%
\url{https://doi.org/10.1145/3369740.3369784}
\showURL{%
\tempurl}


\bibitem[\protect\citeauthoryear{Kashyap, Calciu, Cheng, Min, and Kim}{Kashyap
  et~al\mbox{.}}{2019}]%
        {sosp19-kashyap}
\bibfield{author}{\bibinfo{person}{Sanidhya Kashyap}, \bibinfo{person}{Irina
  Calciu}, \bibinfo{person}{Xiaohe Cheng}, \bibinfo{person}{Changwoo Min},
  {and} \bibinfo{person}{Taesoo Kim}.} \bibinfo{year}{2019}\natexlab{}.
\newblock \showarticletitle{Scalable and Practical Locking with Shuffling}. In
  \bibinfo{booktitle}{\emph{Proceedings of the 27th ACM Symposium on Operating
  Systems Principles}} \emph{(\bibinfo{series}{SOSP ’19})}.
  \bibinfo{publisher}{Association for Computing Machinery}.
\newblock
\showISBNx{9781450368735}
\urldef\tempurl%
\url{https://doi.org/10.1145/3341301.3359629}
\showURL{%
\tempurl}


\bibitem[\protect\citeauthoryear{Kashyap, Min, and Kim}{Kashyap
  et~al\mbox{.}}{2017}]%
        {usenixatc17-kashyap}
\bibfield{author}{\bibinfo{person}{Sanidhya Kashyap}, \bibinfo{person}{Changwoo
  Min}, {and} \bibinfo{person}{Taesoo Kim}.} \bibinfo{year}{2017}\natexlab{}.
\newblock \showarticletitle{Scalable NUMA-aware Blocking Synchronization
  Primitives}. In \bibinfo{booktitle}{\emph{2017 {USENIX} Annual Technical
  Conference ({USENIX} {ATC} 17)}}. \bibinfo{publisher}{{USENIX} Association}.
\newblock
\urldef\tempurl%
\url{https://www.usenix.org/conference/atc17/technical-sessions/presentation/kashyap}
\showURL{%
\tempurl}


\bibitem[\protect\citeauthoryear{Lim and Agarwal}{Lim and Agarwal}{1994}]%
        {asplos94-lim}
\bibfield{author}{\bibinfo{person}{Beng-Hong Lim} {and} \bibinfo{person}{Anant
  Agarwal}.} \bibinfo{year}{1994}\natexlab{}.
\newblock \showarticletitle{Reactive Synchronization Algorithms for
  Multiprocessors}. In \bibinfo{booktitle}{\emph{Proceedings of the Sixth
  International Conference on Architectural Support for Programming Languages
  and Operating Systems}} \emph{(\bibinfo{series}{ASPLOS VI})}.
  \bibinfo{publisher}{ACM}.
\newblock
\urldef\tempurl%
\url{https://doi.org/10.1145/195473.195490}
\showDOI{\tempurl}


\bibitem[\protect\citeauthoryear{Long}{Long}{2013}]%
        {Long13}
\bibfield{author}{\bibinfo{person}{Waiman Long}.}
  \bibinfo{year}{2013}\natexlab{}.
\newblock \bibinfo{title}{qspinlock: Introducing a 4-byte queue spinlock
  implementation}.
\newblock \bibinfo{howpublished}{\url{https://lwn.net/Articles/561775}, July
  31, 2013}.
\newblock
\newblock
\shownote{Accessed: 2018-09-19.}


\bibitem[\protect\citeauthoryear{Luchangco, Nussbaum, and Shavit}{Luchangco
  et~al\mbox{.}}{2006}]%
        {europar06-luchangco}
\bibfield{author}{\bibinfo{person}{Victor Luchangco}, \bibinfo{person}{Dan
  Nussbaum}, {and} \bibinfo{person}{Nir Shavit}.}
  \bibinfo{year}{2006}\natexlab{}.
\newblock \showarticletitle{Hierarchical CLH Queue Lock}. In
  \bibinfo{booktitle}{\emph{Euro-Par 2006 Parallel Processing}}.
  \bibinfo{publisher}{Springer Berlin Heidelberg}.
\newblock
\urldef\tempurl%
\url{https://doi.org/10.1007/11823285_84}
\showURL{%
\tempurl}


\bibitem[\protect\citeauthoryear{M.~Auslander and Wisniewski}{M.~Auslander and
  Wisniewski}{2003}]%
        {K42}
\bibfield{author}{\bibinfo{person}{O.~Krieger B.~Rosenburg M.~Auslander,
  D.~Edelsohn} {and} \bibinfo{person}{R. Wisniewski}.}
  \bibinfo{year}{2003}\natexlab{}.
\newblock \bibinfo{title}{Enhancement to the MCS lock for increased
  functionality and improved programmability -- U.S. patent application number
  20030200457}.
\newblock
\newblock
\urldef\tempurl%
\url{https://patents.google.com/patent/US20030200457}
\showURL{%
\tempurl}


\bibitem[\protect\citeauthoryear{{Magnusson}, {Landin}, and
  {Hagersten}}{{Magnusson} et~al\mbox{.}}{1994}]%
        {ipss94-magnusson}
\bibfield{author}{\bibinfo{person}{P. {Magnusson}}, \bibinfo{person}{A.
  {Landin}}, {and} \bibinfo{person}{E. {Hagersten}}.}
  \bibinfo{year}{1994}\natexlab{}.
\newblock \showarticletitle{Queue locks on cache coherent multiprocessors}. In
  \bibinfo{booktitle}{\emph{Proceedings of 8th International Parallel
  Processing Symposium}}.
\newblock
\urldef\tempurl%
\url{https://doi.org/10.1109/IPPS.1994.288305}
\showDOI{\tempurl}


\bibitem[\protect\citeauthoryear{Mellor-Crummey and Scott}{Mellor-Crummey and
  Scott}{1991}]%
        {tocs91-MellorCrummey}
\bibfield{author}{\bibinfo{person}{John~M. Mellor-Crummey} {and}
  \bibinfo{person}{Michael~L. Scott}.} \bibinfo{year}{1991}\natexlab{}.
\newblock \showarticletitle{Algorithms for Scalable Synchronization on
  Shared-memory Multiprocessors}.
\newblock \bibinfo{journal}{\emph{ACM Trans. Comput. Syst.}}
  (\bibinfo{year}{1991}).
\newblock
\urldef\tempurl%
\url{http://doi.acm.org/10.1145/103727.103729}
\showURL{%
\tempurl}


\bibitem[\protect\citeauthoryear{Radovi{\'c} and Hagersten}{Radovi{\'c} and
  Hagersten}{2003}]%
        {HBO}
\bibfield{author}{\bibinfo{person}{Zoran Radovi{\'c}} {and}
  \bibinfo{person}{Erik Hagersten}.} \bibinfo{year}{2003}\natexlab{}.
\newblock \showarticletitle{{Hierarchical Backoff Locks for Nonuniform
  Communication Architectures}}. In \bibinfo{booktitle}{\emph{International
  Symposium on High Performance Computer Architecture -- HPCA}}.
  \bibinfo{publisher}{IEEE Computer Society}.
\newblock
\urldef\tempurl%
\url{http://dl.acm.org/citation.cfm?id=822080.822810}
\showURL{%
\tempurl}


\bibitem[\protect\citeauthoryear{{Schweizer}, {Besta}, and
  {Hoefler}}{{Schweizer} et~al\mbox{.}}{2015}]%
        {pact15-schweizer}
\bibfield{author}{\bibinfo{person}{H. {Schweizer}}, \bibinfo{person}{M.
  {Besta}}, {and} \bibinfo{person}{T. {Hoefler}}.}
  \bibinfo{year}{2015}\natexlab{}.
\newblock \showarticletitle{Evaluating the Cost of Atomic Operations on Modern
  Architectures}. In \bibinfo{booktitle}{\emph{2015 International Conference on
  Parallel Architecture and Compilation (PACT)}}.
\newblock
\urldef\tempurl%
\url{https://doi.org/10.1109/PACT.2015.24}
\showDOI{\tempurl}


\bibitem[\protect\citeauthoryear{Scott}{Scott}{2013}]%
        {Scott2013}
\bibfield{author}{\bibinfo{person}{Michael~L. Scott}.}
  \bibinfo{year}{2013}\natexlab{}.
\newblock \bibinfo{booktitle}{\emph{Shared-Memory Synchronization}}.
\newblock \bibinfo{publisher}{Morgan \& Claypool Publishers}.
\newblock
\showISBNx{160845956X, 9781608459568}


\bibitem[\protect\citeauthoryear{Theil}{Theil}{1967}]%
        {Theil}
\bibfield{author}{\bibinfo{person}{H. Theil}.} \bibinfo{year}{1967}\natexlab{}.
\newblock \bibinfo{booktitle}{\emph{Economics and Information Theory}}.
\newblock \bibinfo{publisher}{North-Holland}.
\newblock


\bibitem[\protect\citeauthoryear{Verner, Mendelson, and Schuster}{Verner
  et~al\mbox{.}}{2017}]%
        {turbo}
\bibfield{author}{\bibinfo{person}{U. Verner}, \bibinfo{person}{A. Mendelson},
  {and} \bibinfo{person}{A. Schuster}.} \bibinfo{year}{2017}\natexlab{}.
\newblock \showarticletitle{Extending Amdahl’s Law for Multicores with Turbo
  Boost}.
\newblock \bibinfo{journal}{\emph{IEEE Computer Architecture Letters}}
  (\bibinfo{year}{2017}).
\newblock
\urldef\tempurl%
\url{https://doi.org/10.1109/LCA.2015.2512982}
\showURL{%
\tempurl}


\bibitem[\protect\citeauthoryear{{Wikipedia Contributors}}{{Wikipedia
  Contributors}}{2020a}]%
        {rstddev}
\bibfield{author}{\bibinfo{person}{{Wikipedia Contributors}}.}
  \bibinfo{year}{2020}\natexlab{a}.
\newblock \bibinfo{title}{Coefficient of Variation}.
\newblock
\newblock
\urldef\tempurl%
\url{https://en.wikipedia.org/wiki/Coefficient_of_variation}
\showURL{%
\tempurl}


\bibitem[\protect\citeauthoryear{{Wikipedia Contributors}}{{Wikipedia
  Contributors}}{2020b}]%
        {MatthewEffect}
\bibfield{author}{\bibinfo{person}{{Wikipedia Contributors}}.}
  \bibinfo{year}{2020}\natexlab{b}.
\newblock \bibinfo{title}{Matthew Effect}.
\newblock
\newblock
\urldef\tempurl%
\url{https://en.wikipedia.org/wiki/Matthew_effect}
\showURL{%
\tempurl}


\bibitem[\protect\citeauthoryear{{Wikipedia Contributors}}{{Wikipedia
  Contributors}}{2020c}]%
        {Theil-w}
\bibfield{author}{\bibinfo{person}{{Wikipedia Contributors}}.}
  \bibinfo{year}{2020}\natexlab{c}.
\newblock \bibinfo{title}{Theil index}.
\newblock
\newblock
\urldef\tempurl%
\url{https://en.wikipedia.org/wiki/Theil_index}
\showURL{%
\tempurl}


\end{thebibliography}




\section{Appendix : Algorithmic and Implementation Variations}

\Invisible{\Boldly{CNA with support for expedited admission : } 
CNA can trivially be extended to better support real-time or FIFO acquisition requests
by marking the MCS queue element as \emph{expedited}.  Such expedited elements are
will not be culled into the secondary list of remote threads.   Thus, if S is
the MCS queue element associated with an expedited lock acquisition operation, 
then no threads that arrived after S will be admitted before S.  

In the context of Fissile locks, once an expedited thread acquires the CNA inner
lock, it can then immediately assert impatience, inhibiting bypass over the outer TS lock,
and providing the desired admission service.  
} 

\Boldly{CNA - triggering flushes: } 
Classic CNA runs Bernoulli trials in the unlock path
to decide whether to flush the remote chain into primary and change
the preferred NUMA node, in order to provide long-term fairness.  
We have experimented with variants where the head of the remote chain monitors 
how long it has waited and, if necessary, sets a flag in its MCS queue element 
to cue flushing the remote list into the primary MCS chain, in order to avoid starvation.   
The CNA unlock operator checks that flag, and if set, flushes the remote queue and
changes the preferred NUMA node.  This approach yields a time-based anti-starvation policy 
instead of the count-based Bernoulli trials as found in the original CNA
and shifts the Bernoulli trial out of the unlock path, replacing it with a fetch
of a location that is usually in cache.  In addition, we can use polite constructs
such as MONITOR-MWAIT for timekeeping.  

\Boldly{Probabilistic bounded bypass :} 
We can provided bounded bypass over the outer TS lock as follows, without
requiring an explicit ``Impatient'' state to be encoded or stored in the lock structure.
Briefly, arriving threads run a biased Bernoulli trial with probability of success $P=1/256$.  
On success, threads immediately divert into the CNA slow-path and skip trying the outer TS
fast-path.  If we have a set of threads dominating the outer TS lock and starving the CNA inner lock
owner, they will eventually self-decimate and pass through the CNA inner lock, providing the 
desired anti-starvation. 
This approach is simple, effective, and avoids state and coherence updates associated with state.

Bypass also ensures a ``trickle'' of threads will pass over the CNA inner lock, acting
to homongenize the set of active circulating threads and reduce lock migration.  

\Invisible{PrbBB; Unlock() is simple ST 0; count-based instead of time-based impatience; 
Allows CNA inner lock to act as a sieve/strainer to filter out lock circulating in ACS
over outer TS lock.  Claim and conjecture: lazy PrbBB filter suffices to eventually
homogenize the ACS; } 

\Boldly{Compact single-word form :} 
We can construct a single-world compact form of Fissile by collapsing the Outer, Inner, and Impatient
fields into a single word.   This condensed version is appropriate to replace the 
linux kernel's \emph{qspinlock} construct \cite{arxiv-TWA,linux-locks,Long13}.  
Briefly, the least significant byte serves as the outer lock, the next most significant bit 
encodes the impatient state, and the remaining higher order bits encode the tail of the 
CNA MCS queue.  The Fissile unlock operator stores 0 into the separately addressable
low-order byte to release the lock.  We note that this encoding requires mixed sized atomic
accesses to the same location, the safety of which are platform-dependent. 
The current qspinlock implementation also depends on mixed size accesses.


\Invisible{claim Compact : collapse inner CNA and TS outer into a single word for use in the linux kernel}

\Boldly{Deferred release of the CNA inner Lock :} 
We have investigated deferring the release of the CNA inner lock until we unlock the Fissile lock proper
and specifically after dropping the outer TS lock.  This may improve scalability by shifting CNA 
administrative work (culling and flushing activities) outside and after the TS critical section.
While appealing, this change means that MCS queue elements can not be allocated on-stack, necessitating
more complex queue element lifecycle memory management.  Each thread must have one allocated queue 
element for each lock that it holds whereas with on-stack allocation, each thread has at most one
active queue element.  Furthermore, we must convey the address of the MCS queue element -- 
the CNA owner's element -- from the Fissile acquire operation to the corresponding unlock operation. 
As noted above, typical locking APIs do not have provisions to pass such information from acquire to unlock. 
A viable solution is to implement a thread-local cache that contains at most one element, a reference
to MCS queue element for the most recently acquired inner lock held that thread.

In the Fissile slow path, either before or after acquiring the CNA inner lock, a thread checks its cache.
If a pending deferred queue element is present, it releases the associated CNA lock and clears the cache.
It can then install the current queue element into its cache.  In the Fissile unlock operation, after
releasing the TS outer lock, the thread again checks and clears its cache, releasing any locks associated
with a pending queue element found in the cache.  

The following benign scenario can arise.
Thread $T1$ acquires contended lock $L1$ via the slow path using MCS queue element $E1$.  
Our approach defers the release of the CNA inner lock of $L1$ and stores $E1$ into $T1$'s cache.  
$T1$ then goes on to acquire another contended lock $L2$ using MCS element $E2$, 
displacing $E1$ from $T1's$ cache and forcing the early release of $L1$'s inner CNA lock 
via $E1$.  $E1$ can then be reclaimed and $E2$ is installed in the cache.  
As noted, the premature release of $L1$'s inner CNA lock is harmless.

\Boldly{Deferred wakeup of the CNA successor :}  
Instead of deferring the release of the CNA inner lock until the Fissile unlock, and passing 
the queue element reference to from the Fissile acquire operation to the corresponding unlock operation, 
we can perform a \emph{partial MCS release} in the Fissile lock operation.  The partial release operator 
clears the MCS \texttt{Tail} variable if there are no other elements on the chain, and otherwise 
identifies the successor -- the next element in the MCS chain.  If there is a successor in the chain, 
the partial release does \emph{not} notify that thread.  Instead, we pass a reference to the successor
queue element to the corresponding Fissile unlock operator, which in turn passes ownership to the 
successor via setting the usual MCS flag in the successor's queue element.  
This again allows the queue elements to be allocated on-stack and simplifies memory management.   

As above, we can employ a thread-local cache of 1 element to help convey the successor reference
to the unlock operator.  The deferred notification can occur at any point between acquiring the
TS outer lock to immediately after releasing the TS outer lock.  Thread $T1$ acquires contended
lock $L1$ via the slow path, having performed a partial release on the inner lock of $L1$, 
identifying thread $T2$ (or more precisely the MCS node associated with $T2$) as the successor.  
$T1$ then sets its thread-local cache to refer to the successor on $L1$, thread $T2$.  
When $T1$ releases $T2$, it notifies $T2$ -- via $T2$'s queue element -- so $T2$ starts running 
as the owner of $L1$'s inner lock.   However $T1$, while holding $L1$, might acquire 
contend lock $L2$ via the slow path.  During acqusition of $L2$, $T1$ will need to displace $T2$ from 
its cache, waking $T2$ early, and installing the successor for $L2$, $T3$, in its cache.  
Such early wakeup is benign.   

\Boldly{Simplified encoding of the ``Impatient'' state :}
We remove the explicit Impatient field from the Fissile lock structure.
The outer lock word encoding changes as follows : 0 indicates unlocked; 1 indicates locked; 2 
indicates locked but with an impatient alpha thread.  
When the alpha thread becomes impatient, it executes an atomic fetch-and-increment
on the outer TS lock word.  If value advanced from 0 to 1, then the alpha thread acquired
the outer TS lock.   Otherwise the value advanced from 1 to 2 indicating TS lock was held 
by some other thread. The alpha thread waits impatiently for the word to change
back to 1, at which point it has gained ownership of the outer TS lock.  
The Fissile unlock operator simply executes an atomic decrement on the lock word
shifting the value from either 2 to 1 or 1 to 0.  
While more compact, this variant requires an atomic decrement in the unlock path, instead 
of a simple store.  We note that the form that uses an explicit store depends on eventual consistency,
where threads in unlock will eventually observe the Impatient = 2 value set by an
impatiently waiting alpha thread.  

\Boldly{3-Stage Gated}
We note that we can use an \emph{3-Stage} construction \cite{arxiv-TWA}.  The embodiment described
above has 2 stages of waiting for threads that encounter contention : the inner lock and the outer lock.  
Adding an extra \emph{Gate} stage can, paradoxically, act to reduce handover latency.  
Contended threads in the slow path proceed as follows : 
acquire the CNA inner lock; wait until the Gate becomes 0; 
set the Gate to 1; release the inner CNA lock; wait for acquire the outer TS lock; clear the Gate; 
execute the critical section; and finally release the outer TS lock.  
At any given time there can be at most one thread waiting on the gate and 
at most one thread busy-waiting on the outer lock.  
Clearing the Gate entails low latency as there is at most one thread waiting on the 
gate, allowing faster handover and improved pipelining of lock acquisition operations.  
Note that acquiring the Gate occurs only while holding the inner lock, so atomics are not required.  

Instead of a simple gate, which allows at most one thread, we could also employ a semaphore to allow 
a very small number of threads to busy-wait on the outer TS lock.  

\Invisible{Flow schematic} 

Slow path flow :
$ [Inner(N)] \mapsto [Gate(1)] \mapsto *Inner \mapsto [Outer(1)] \mapsto *Gate \mapsto CS \mapsto *Outer \mapsto NCS  $
; where $[Inner(N)]$ indicates that at most $N$ threads wait to acquire the inner lock, and $*Inner$ reflects
the corresponding release of the inner lock.

\Boldly{3-Stage with outer ticket lock} 
Borrowing from \emph{TWA} \cite{arxiv-TWA} (\emph{TWA-Staged and 3-Stage} variations) we can easily 
construct a 3-Stage Fissile lock where the outer lock is a ticket lock, with 
differentiated \emph{near} and \emph{far} waiting on the ticket lock. 
Threads in the slow path proceed as follows: acquire the CNA inner lock; Fetch and increment the ticket variable
to assign a unique ticket value to the locking request; busy-wait while the assigned ticket value differs from 
the ticket lock's grant field by 2 or more (\emph{far} waiting); release the CNA inner lock; busy-wait until
the assigned ticket equals the grant variable (\emph{near} waiting); 
enter and execute the critical section; release the outer ticket lock by incrementing the grant variable.  
A non-atomic increment suffices in the unlock path.  
In this formulation no bypass is allowed, although a fast path is feasible.  The CNA inner lock
completely dictates the order of admission, and we use the ticket locks, which have efficient
handover under light contention, when a thread nears the front of the conceptual queue of waiting threads.
A very small number of threads wait on the ticket lock at any given time, leveraging its 
excellent behavior in that mode.  

\Boldly{Impatience policies:} 
Myriad policies are possible for setting and clearing the ``impatient'' state.  
Our reference implementation uses a simple form where the alpha thread waits, if
necessary becomes impatient, and then cancels the impatient state once it obtains
the outer TS lock.   Other possibilities include leaving the impatient state set
for all threads currently waiting on the inner lock, or until the inner 
lock drains to empty state \cite{GoMutex}. 

\Invisible{The ``Go'' Mutex uses a mostly LIFO semaphore in a manner
similar to that of the Fissile inner lock, and similar to the \emph{LIFO-CR} construct
described in \cite{arxiv-Malthusian}.}  

\Boldly{TS Tunables:} 
Our current implement allows arriving thread to try the TS just once.  
Experiments suggest, however, some benefit to allowing a brief bounded 
polite spinning phase, potentially with back-off, before diverting into the slow-path, 
as long as the lock remains impatient.  In addition, in our current implementation, 
the alpha thread uses a simple busy-wait loop with no back-off.  
Employing some moderate back-off policy may be useful.     
While these avenues show promise, our current implementation has no TTS tunable
values and exhibits desirable \emph{parameter parsimony}.  

\Boldly{Spin-then-Park waiting:}
As described, stalled threads use simple busy-waiting.  We note, however,
that it relatively simple to modify Fissile locks so that threads waiting
on the CNA inner lock will \emph{park} -- descheduling themselves via the operating
system so the CPU where they were running can dispatch other ready threads or
become idle.

\section{Appendix: Remarks} 

\Boldly{CNA Administrative responsibilities:} 
\\
NUMA-aware reorganization of the MCS chain : Who; what; where; when; how.  In order of preference:
\begin{enumerate}[leftmargin=*,align=left] 
\item A waiter on the chain reorganizes outside the critical section -- delegated helping.  
\item A thread reorganizes the chain after having dropped the lock, outside and after the CS.
This thread could still be performing useful work, however, so we are borrowing it to help.
\item Threads reorganize the chain in unlock while still holding the lock, on
the critical path,  potentially extending the effective critical section duration.
This impacts lock hand-over response time -- the time needed to convey ownership to a successor --
and scalability.  
Classic CNA-MCS uses this approach. 
\end{enumerate} 

Shuffle, for instance, delegates NUMA reorganization to other waiting threads, 
allow parallelism between reorganization and execution of the critical section. 
While elegant, for short critical sections with intense contention the
coherent communications cost can come to dominate performance, and make this approach unprofitable.
That is, the granularity of work being delegated does not overcome the communication costs to
delegate and coordinate.    
And when the critical sections are longer then any additional CNA overheads in the
critical path are less important, so delegated execution often has no appreciable benefit
relative to CNA.  


\Boldly{TS : polite vs impolite}
Anderson \cite{tpds90-Anderson} observed that ``polite'' test-and-test-and-set locks
may be a better choice for contended locks than simple ``impolite'' test-and-set locks.
Test-and-test-and-set locks are polite in the sense that they first load and check
the lock word before conditionally attempting the atomic operation to acquire the lock. 
Simple test-and-set locks are optimistic and forego that load and check and simply 
try the atomic operation.  
But if the lock is already held, such futile atomic operators may generate unnecessary
coherence and write invalidation.  A test-and-test-and-set strategy acts to reduce 
those overheads and the the rate of failed atomic operations.  

We have found, however, that a simple impolite impolite test-and-set policy
is appropriate for the outer Fissile TS lock, ostensibly as there is at most 
one thread busying waiting on the outer lock at any given time.  
Relatedly, polite test-and-test-and-set TTS acquisition may incur more coherence 
bus transactions in the case where the lock in not held but the lock word is not in cache or 
is in dirty state in a remote cache, as would be the case if the previous owner
ran on a CPU with a different cache.  Absent coherence probe speculation, on 
bus operation is needed to load the lock and a second to upgrade the cache line
to written state.  This scenario is common if the lock instance is \emph{promiscuous} 
-- being lock by multiple different threads, but with little or no contention.  

Performant test-and-set locks will typically probe the lock directly with an atomic
on arrival, optimistically assuming they can acquire the lock, but then shift
to a polite TTS mode as they busy wait.  This strategy reduces latency in the
uncontended case but acts to unnecessary write invalidation and coherence
traffic in the contended case.  

\Invisible{Polite lock attempts avoid futile atomic operators which generate 
unnecessary coherence traffic via write invalidation if the lock is found held.} 

\Boldly{Anti-starvation:} 
Fissile requires two types of anti-starvation. The first is in CNA, to ensure that
lock operations from remote NUMA nodes are eventually serviced, and the second is managed
by the alpha thread to avoid indefinite bypass over the TS outer lock. 

\Boldly{Linux kernel qspinlock:} 

The Linux \emph{qspinlock} construct \cite{linux-locks,linux-page-struct,Long13} is a 
compact 32-bit lock, even on 64-bit architectures.  The low-order bits of the lock word 
constitute a simple test-and-set lock while the upper bits encode the tail of an MCS chain.  
The result is a hybrid of MCS and test-and-set\footnote{\url{https://github.com/torvalds/linux/blob/master/kernel/locking/qspinlock.c}}. 
In order to fit into a 32-bit work -- a critical requirement -- the chain is formed by 
logical CPU identifiers instead of traditional MCS queue node pointers.  
Arriving threads attempt to acquire the test-and-set lock embedded in the low order bits 
of the lock word.  This attempt fails if the test-and-set lock is held or of the 
MCS chain is populated.  If successful, they enter the critical section, otherwise 
they join the MCS chain embedded in the upper bits of the lock word.  When a 
thread becomes an owner of the MCS lock, it can wait for the test-and-set lock 
to become clear, at which point it claims the test-and-set lock, releases the MCS lock, 
and then enters the critical section.  The MCS aspect of qspinlock is used only 
when there is contention.  The unlock operator simply clears the test-and-set lock.  
The MCS lock is never held over the critical section, but only during contended 
acquisition.  Only the owner of the MCS lock spins on the test-and-set lock,
reducing coherence traffic \footnote{This provides a LOITER-style \cite{eurosys17-dice} 
lock with the \emph{outer lock} consisting of a test-and-set lock and the \emph{inner lock} 
consisting of the MCS lock, with both locks embedded in the same 32-bit word.}.  
Qspinlock is strictly FIFO.  While the technique employs local spinning on the MCS chain, 
unlike traditional MCS, arriving and departing threads will both update the common lock word, 
increasing coherence traffic and degrading performance relative to classic MCS.   
Qspinlock incorporates an additional optimization where the first contending thread 
spins on the test-and-set lock instead of using the MCS path.  
Traditional MCS does not fit well in the Linux kernel as (a) the constraint 
that a low-level spin lock instance be only 32-bits is a firm requirement, and (b) 
the lock-unlock API does not provide a convenient way to pass the owner's MCS queue 
node address from lock to unlock.  We note that qspinlocks replaced classic ticket 
locks as the kernel's primary low-level spin lock mechanism in 2014, and ticket 
locks replaced test-and-set locks, which are unfair and allow unbounded bypass, 
in 2008 \cite{ticket-spinlocks}.   

We note that the kernel provides a specialized qspinlock form for
paravirtualized environments within virual machines, the so-called ``PV-friendly" qspinlock.
We believe the same properties that make Fissile tolerant of preemption 
also make it inherently PV-friendly, and, if used in the kernel, Fissile may obviate
the need for a different PV-friendly form.  

\Boldly{Original abstract} 
Classic \emph{test-and-test} (TS) mutual exclusion locks \cite{tpds90-Anderson}
are simple, enjoy high performance and low latency of ownership transfer under
light or no contention but do not, however, scale gracefully
under high contention. Furthermore TS locks do not provide any admission order
guarantees, and may allow sustained starvation of waiting threads and
long-term unfairness.

Such concerns led to the development of scalable queue-based locks
such as \emph{MCS locks} (Mellor-Crummey and Scott) \cite{tocs91-MellorCrummey} 
and NUMA-aware variants thereof such
as \emph{Compact NUMA-aware Locks} (CNA) \cite{EuroSys19-CNA, arxiv-CNA}.  
Both MCS and CNA scale under load, but have more complicated lock handover operations
than TS and suffer higher latencies at low contention. 

We propose \underline{\textbf{Fissile}} locks, which capture the most desirable properties
of both TS and CNA.  A Fissile lock consists of two underlying locks: a TS lock and a CNA lock.  
Acquiring ownership of the TS lock confers ownership of the compound Fissile Lock. 
Arriving threads first use an atomic instruction to try to acquire the TS lock. 
If successful, they immediately enter the critical section, and we say the
Fissile lock was acquired via the \emph{fast path}. 
If the fast path attempt fails, the thread then acquires the CNA lock and
busy-waits on the TS lock, releases the CNA lock, and enters the critical section.
Releasing a Fissile lock entails just releasing the TS lock.  
Contended locking uses the CNA lock while uncontended operations use the TS lock. 
Fissile locks deflect contention away from TS lock into the CNA lock. 
 
To avoid TS-based starvation, the thread holding the CNA lock and waiting on the 
TS lock can become ``impatient'' and cue direct handover of ownership the next time the 
TS lock is released, bounding bypass.  The result is a highly scalable NUMA-aware lock 
that performs like TS at low contention, enjoying low latency, and like CNA at high contention.   

\Boldly{Comparison of lock properties: }

\begin{table} [h]
\centering
{\fontsize{6.5}{6.5}\selectfont
\begin{tabular}{lcccc}
\toprule
\cmidrule(lr){2-3}
\cmidrule(lr){4-5}

&
\multicolumn{1}{c}{NUMA-Aware} &
\multicolumn{1}{c}{Bypass} &
\multicolumn{1}{c}{TS Fast-path} &
\multicolumn{1}{c}{Uncontended Unlock} \\
\midrule

QSpinlock      & No    & No      & Yes  & Store    \\
Go Mutex       & No    & Bounded & Yes  & Atomic Decrement  \\
MCS            & No    & No      & No   & CAS       \\ 
CNA            & Yes   & No      & No   & CAS       \\
QSpinlock+CNA  & Yes   & No      & Yes  & Store     \\
Shuffle Locks  & Yes   & No      & Yes  & Store     \\ 
Fissile Locks  & Yes   & Bounded & Yes  & Store     \\

\midrule[\heavyrulewidth]
\bottomrule
\end{tabular}%
}
\caption{Lock Properties}\label{Table:LockProperties}
\end{table}

\end{document}